\documentclass[12pt]{iopart}
\usepackage{graphicx}
\usepackage{iopams}
\usepackage{setstack}
\usepackage{array}
\usepackage{amstext}
\renewcommand{\theenumi}{\alph{enumi}}
\begin{document}
\title{A Dynamic Programming Approach to Adaptive Fractionation}
\author{Jagdish Ramakrishnan$^1$, David Craft$^2$, Thomas Bortfeld$^2$, and John N. Tsitsiklis$^1$}
\address{$^1$ Laboratory for Information and Decision Systems, Massachusetts Institute of Technology, Cambridge, MA 02139, USA \\ $^2$ Department of Radiation Oncology, Massachusetts General Hospital and Harvard Medical School, Boston, MA 02114, USA}
\ead{jagdish@mit.edu, dcraft@partners.org, tbortfeld@partners.org, and jnt@mit.edu}
\date{September 5, 2011}
\begin{abstract}
% Previous: We formulate a previously introduced adaptive fractionation problem in a dynamic programming (DP) framework and explore various solution techniques. The two messages of this paper are: (i) the DP model is a useful framework for studying adaptive radiation therapy, particularly adaptive fractionation, and (ii) there is a potential for substantial decrease in dose to the primary organ-at-risk (OAR), or equivalently increase in tumor escalation, when using an adaptive fraction size. 
We conduct a theoretical study of various solution methods for the adaptive fractionation problem. The two messages of this paper are: (i) dynamic programming (DP) is a useful framework for adaptive radiation therapy, particularly adaptive fractionation, because it allows us to assess how close to optimal different methods are, and (ii) heuristic methods proposed in this paper are near-optimal, and therefore, can be used to evaluate the best possible benefit of using an adaptive fraction size. 

The essence of adaptive fractionation is to increase the fraction size when the tumor and organ-at-risk (OAR) are far apart (a ``favorable" anatomy) and to decrease the fraction size when they are close together. Given that a fixed prescribed dose must be delivered to the tumor over the course of the treatment, such an approach results in a lower cumulative dose to the OAR when compared to that resulting from standard fractionation. We first establish a benchmark by using the DP algorithm to solve the problem exactly. In this case, we characterize the structure of an optimal policy, which provides guidance for our choice of heuristics. We develop two intuitive, numerically near-optimal heuristic policies, which could be used for more complex, high-dimensional problems. Furthermore, one of the heuristics requires only a statistic of the motion probability distribution, making it a reasonable method for use in a realistic setting. Numerically, we find that the amount of decrease in dose to the OAR can vary significantly (5 - 85\%) depending on the amount of motion in the anatomy, the number of fractions, and the range of fraction sizes allowed. In general, the decrease in dose to the OAR is more pronounced when: (i) we have a high probability of large tumor-OAR distances, (ii) we use many fractions (as in a hyper-fractionated setting), and (iii) we allow large daily fraction size deviations.
\end{abstract}
%\submitto{\PMB}
\maketitle

\section{Introduction}

In its broadest context, adaptive radiation therapy (ART) is a radiation treatment process that uses feedback information to modify and improve treatment plans \cite{YVW1997, Lan2011}. Feedback information could include patient anatomy information such as positions of tumor and organ-at-risk (OAR), and can be obtained from imaging modalities such as cone-beam computed tomography (CBCT), ultrasound imaging, or portal imaging \cite{PTE2010}. 

We can correct for inter-fractional variations in patient anatomy by adapting a treatment plan either off-line or on-line. An off-line adaptation uses imaging information available after the delivery of a fraction to modify the treatment plan for the next fraction. On the other hand, an on-line adaptation uses information acquired immediately before the delivery of a fraction for a quick modification of the treatment plan for that fraction. The advantage of on-line ART is the availability of more data (inclusion of patient anatomy for the current fraction). However, due to patient wait time and treatment duration limitations, on-line ART requires (i) fast re-optimization of the treatment plan, and (ii) re-planning across a small number of degrees of freedom. Conversely, in off-line ART, a re-optimized treatment plan can be determined on a slower time-scale. 

Due to the immediate possibility of lower cumulative dose to healthy organs through treatment plan re-optimization between fractions, off-line ART has received much attention in the research community. One of the early approaches involved using information about tumor variations (both systematic and random) during the first few fractions to determine a customized treatment plan for the remaining fractions \cite{YVW1997, YZJ1998}. The customized treatment plan generally has a smaller planning target volume (PTV) suited to the particular patient. Such adaptation is shown to improve treatment efficacy and to allow for dose escalation to the tumor \cite{YLB2000, GWR2011}. Another approach focused on using a smaller PTV initially and re-optimizing treatment plans between fractions to compensate for the accumulated dose errors \cite{RFL2004, ZAX2007, DeF2008, Web2008a, WeB2008b}. We do note that the practical applicability of this method relies on the ability to accurately determine the delivered dose at each voxel. However, determining the delivered dose accurately requires reliable deformable registration algorithms, which is still a major research topic. 

In on-line ART, the focus has been on making adjustments to the existing treatment plan rather than on re-optimizing for an entirely new plan. This is primarily because the time between the acquisition of patient anatomy information and the delivery of a plan is on the order of minutes rather than hours. In this case, a full re-optimization and complete quality assurance of the treatment plan may not be possible. Several on-line ART approaches have been developed which make quick modifications to either the fluence map or multi-leaf collimator (MLC) leaves to match the planned dose \cite{MZW2005, CDL2005, WTW2008}. Whereas these methods involve spatially varying the dose distribution, other methods, including the work in this paper, consider temporally varying the fraction size from day to day \cite{LCC2008, CLC2008}. 

\begin{figure}
\centerline{\includegraphics[width=120mm]{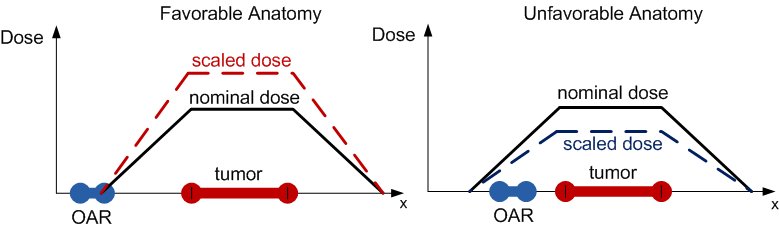}}
% added parenthetical descriptions -- when the tumor and OAR are far apart and close together, respectively.
\caption{Adaptive fractionation capitalizes on tumor-OAR variations. Nominal dose corresponds to leaving the fraction size unchanged, while scaled dose corresponds to a changed fraction size. When we have a favorable anatomy (i.e., the tumor and OAR are far apart) as in the left panel, we can use a larger fraction size. Similarly, for an unfavorable anatomy (i.e., the tumor and OAR are close together) as in the right panel, we can use a smaller fraction size. Our model is more general than this 1-dimensional example and can be used for 3-dimensional realistic settings as well.}
\label{fig:dose_scaling_nice}
\end{figure}

% previous first sentence: We discuss the adaptive fractionation problem first introduced in \cite{LCC2008}.
% 
We now briefly motivate the adaptive fractionation problem introduced in \cite{LCC2008}. We focus on a model of the variations of the tumor and one primary OAR, which is usually the limiting factor in escalating the dose to the tumor. Using an adaptive fraction size can allow us to take advantage of a ``favorable" patient anatomy by increasing the fraction size. Similarly, we can decrease the fraction size for an ``unfavorable" anatomy. One simple way to think about this problem is to consider variations of the distance between the tumor and the OAR from day to day (see Figure \ref{fig:dose_scaling_nice}). If the distance is large, we can escalate the dose to the tumor (since the OAR dose per unit tumor dose is small) and vice versa, if the distance is small. Given that a fixed prescribed dose must be delivered to the tumor, adaptive fractionation results in a lower cumulative dose to the OAR over the course of the treatment. We emphasize that our model is more general than this 1-dimensional distance setting and can be applied to 3-dimensional realistic settings as well.

% added reference \cite{Ber2007} for "the DP approach is useful
% removed We believe that, though it is not new to the adaptive therapy field, the DP approach has a good potential for improving the dosimetric quality of a treatment plan.
% major change to sentences of this paragraph
% removed In the adaptive fractionation setting considered in this paper, the algorithm in \cite{LCC2008} can be understood as a variant of the open-loop feedback control approximate DP approach (see \cite{Ber2007} for a description of such an approach).
The purpose of this study is to develop and evaluate solution methods for the adaptive fractionation problem. We use the dynamic programming (DP) algorithm to solve the problem exactly and to assess how close to optimal various heuristic methods are. The DP approach is useful for sequential decision making problems, especially when there is a need for balancing the immediate and future costs associated with making a decision in any particular stage \cite{Ber2007}. For off-line ART, the DP approach can be used to compensate for past accumulated errors in dose to the tumor \cite{FeV2004,DeF2008,SEP2010}. For on-line ART, an approach for adaptive fractionation based on biological models also makes use of DP \cite{CLC2008}. The results of our study indicate that heuristic methods, both the ones proposed in this paper and in \cite{LCC2008}, are near-optimal under most conditions. The consequence is that these methods can be used to evaluate the best possible benefit of using an adaptive fraction size. Furthermore, simple heuristics as proposed in this paper provide a quick way to measure the gain that results from adaptively varying the fraction size. 

% added para:
In Section 2, we formulate the adaptive fractionation problem and describe solution methods in detail. Results from numerical simulations are evaluated in Section 3. Finally, Section 4 includes discussions about realistic implementations, model assumptions, and future directions.

\section{Materials and Methods} 
We now describe the details of the model and formulate the adaptive fractionation problem. Let $N$ be the number of fractions and $P$ be the total prescribed dose to the tumor. The patient anatomy in the $k$th day is represented by a variable denoted by $s_k$, which is sampled, independent of other days, from a known probability distribution $p(\cdot)$ estimated from historical data. We assume that the patient anatomy $s_k$ is observed just before the delivery of the $k$th fraction and can be obtained, for example, from imaging modalities such as CBCT. We could also use this formulation for intra-fractional variations, where $s_k$ would change during a fraction; in this case, it would be more appropriate to assume that patient anatomy instances are correlated rather than independent from one another. In general, the distribution $p(\cdot)$ can be either continuous or discrete but for simplicity, we assume a continuous distribution and denote by $S$ the set of possible anatomy instances. We define $r_k$ to be the remaining prescribed dose left to deliver to the tumor in the $k$th and future fractions. We must determine the fraction size $u_k$ in the $k$th fraction based on the remaining dose $r_k$ and patient anatomy $s_k$. Here, $r_k$ and $s_k$ together represent the state of the system because they are the only relevant pieces of information needed to determine the fraction size $u_k$. It can be seen that the dynamics of the system are described by the equations $r_{k+1} = r_{k} - u_{k}$, $s_k\sim p(\cdot)$, for $k = 1, 2, \ldots, N$, with $r_{1}$ initialized to the prescribed dose $P$.

Given a patient anatomy $s_k$, the dose delivered to the OAR in the $k$th fraction can be written as $u_kh(s_k)$, where $h(s_k)$ is the OAR dose per unit Gy dose delivered to the tumor. For the 1-dimensional setting in Figure \ref{fig:dose_scaling_nice}, the function $h(s_k)$ describes the dose falloff, as a function of the location of the OAR. We could use other choices for $h(s_k)$; what we need is a function that describes how favorable a particular patient anatomy $s_k$ is. If the current technology available allows for a quick way to supply information about the favorability of a patient anatomy before the delivery of a fraction, this information would be captured in the $h(s_k)$ function. For notational convenience, we also define the cumulative distribution function (CDF) for $h(s_k)$ as 
\begin{equation}
F^{\text{h}}(z) = \int_{\{s \, : \, h(s) \leq z \}} \, p(s) \, \rmd s.
\end{equation}

The optimization problem of interest is to minimize the expected total dose to the OAR subject to constraints that ensure that: (i) the prescribed dose to the tumor is met with certainty, and (ii) the fraction size for each day is within a lower bound, $\underline{u}$, and an upper bound, $\overline{u}$. Although optimizing non-linear TCP/NTCP functions would be better choice here, it is simpler to use total dose and is a reasonable surrogate for most situations. For convenience in analysis, we also incorporate constraint (i) into the objective cost function by adding a terminal cost $g(r_{N+1})$, which assigns an infinite penalty when the prescribed dose $P$ is not met. Mathematically, we can formulate the adaptive fractionation problem as follows:
\begin{equation}
\begin{array}{rl}
	 \underset{\{ \mu_{k} \}}{\text{min}}
	& \mathbb{E} \left[ g(r_{N+1}) + \displaystyle\sum_{k = 1}^{N} \mu_k(r_k, s_k) h(s_k) \right] \\ \\
	 \text{s.t.}
	& \underline{u} \leq \mu_{k}(r_{k}, s_{k}) \leq \overline{u}, \qquad \quad \ k = 1, 2, \ldots, N,\quad \forall r_k, \forall s_k \\
	 & r_{1} = P, \\
	 & r_{k+1} = r_{k} - \mu_{k}(r_{k}, s_{k}), \quad \;\;\;\, k = 1, 2, \ldots, N, \\
	 & s_k\sim p(\cdot), \qquad \qquad \qquad \quad \, k = 1, 2, \ldots, N, \\
\end{array} \label{opt_prob}
\end{equation}
where 
\begin{equation}
	g(r_{N+1}) = \left\{
	\begin{array}{lll}
		0,  					&	&\text{if $r_{N+1} = 0$},\\
		\infty,			&	&\text{otherwise},
	\end{array} \right.
	%\label{terminal_cost}
\end{equation}
and where the expectation $\mathbb{E}[\, \cdot \, ]$ is taken with respect to the probability distribution $p(\cdot)$.

In the above optimization problem, we are searching for an \emph{optimal policy} $\mu_{k}^{*}(r_{k}, s_{k})$, which for any given time $k$, is a function of the remaining dose $r_{k}$ and the patient anatomy $s_{k}$. Here, the solution is not simply a single value of the optimal fraction size for any particular day but rather, a policy or a strategy that can possibly choose different fraction sizes based on state information. This is characteristic of closed-loop control, which uses state (feedback) information to make decisions. Furthermore, a brute search over all possible sets of functions $\{ \mu(r_k, s_k) \}$ to solve this problem is not feasible. Notice that the first term in the objective function $g(r_{N+1})$ simply ensures that after $N$ fractions, the prescribed dose to the tumor is met \emph{exactly}. The second term is the total dose to the OAR resulting from using the policy $\mu_k$.

\subsection{A Dynamic Programming (DP) Approach}

We can solve the problem (\ref{opt_prob}) exactly by using the DP algorithm (Bellman's backward recursion):
\begin{equation}
	J_{N+1}(r_{N+1}, s_{N+1}) = g(r_{N+1}) = \left\{
	\begin{array}{lll}
		0,  					&	&\text{if $r_{N+1} = 0$},\\
		\infty,			&	&\text{otherwise},
	\end{array} \right.
	\label{J_N}
\end{equation}
\begin{equation}
J_{k}(r_{k}, s_{k}) = \underset{\underline{u} \leq u_{k} \leq \overline{u}}{\text{min}} \Big( u_{k}h(s_{k}) + \mathbb{E} \left[ J_{k+1}(r_k-u_k, s_{k+1}) \right]\Big), \label{DP_Eq}
\end{equation}
for $k = N, N-1, \ldots, 1$, where the expectation is taken with respect to the distribution $p(\cdot)$ of $s_{k+1}$:
\begin{equation}
\mathbb{E} \left[ J_{k+1}(r_k-u_k, s_{k+1}) \right] = \int_{S} \, p(s) J_{k+1}(r_{k}-u_{k}, s) \, \rmd s. 
\end{equation}
For numerical implementation, however, we need to discretize the variables $r_k$ and $s_k$ and solve a corresponding discrete problem. All of our subsequent results refer to this discrete problem. 

It can be shown that the policy resulting from the above DP algorithm is optimal for the problem (\ref{opt_prob}). Hence, the cost-to-go function $J_{k}(r_k, s_k)$ is the resulting cost from using an optimal policy starting with a given remaining dose $r_{k}$ and patient anatomy $s_{k}$ in the $k$th fraction. The basic idea of the DP algorithm is to start from the last fraction (when the optimal decision $\mu_{N}^*$ must be exactly equal to the remaining dose $r_{N}$ for any possible $s_{N}$), determine the optimal decision $\mu_{N-1}^*$ given this new information, and proceed backwards in determining the present optimal policy with information about future optimal policies. We can see that $J_{k}(r_k, s_k)$ is computed by minimizing the sum of the present cost associated with delivering the fraction size $u_{k}$ in the $k$th fraction (i.e., $u_{k}h(s_{k})$) and the expected future cost resulting from delivering the fraction size $u_{k}$ given that we use an optimal policy thereafter (i.e., $\mathbb{E} \left[ J_{k+1}(r_{k}-u_{k}, s) \right]$). Essentially, the DP algorithm involves precomputing and storing the fraction size $u_k$ off-line for every possible value of the state $(r_k, s_k)$ and fraction $k$. Therefore, choosing the fraction size on-line right before the delivery of the fraction simply involves a quick table lookup. 

We now discuss interesting theoretical properties of an optimal policy resulting from the qualitative structure of the cost-to-function. The piecewise linear structure of the cost-to-go function (see Appendix) results in a special structure of an optimal policy. Essentially, if it is possible to deliver the treatment with a sequence of smallest fraction sizes $\underline{u}$ and largest ones $\overline{u}$, an optimal policy does \emph{exactly} that, i.e., the resulting optimal policy has the threshold structure:
\begin{equation}
	\mu_{k}^*(r_{k}, s_{k}) = \left\{
	\begin{array}{lll}
		\underline{u},		&	&\text{if $h(s_{k}) \geq T_{k}(r_{k})$,}\\
		\overline{u},		&	&\text{if $h(s_{k}) < T_{k}(r_{k})$,}\\
	\end{array} \right. 
	\label{thshd_pol}
\end{equation}
for $k = 1, 2, \ldots, N$, where the $T_k(r_k)$ are pre-computed thresholds (see Appendix for details). The optimal policy (\ref{thshd_pol}) makes sense because unfavorable or large values of $h(s_k)$ result in delivering a small fraction size $\underline{u}$ and vice versa. The policy is completely characterized by the thresholds $T_k(r_k)$, $k = 1, 2, \ldots, N$, which represent the point at which it is optimal to deliver $\underline{u}$ when above it and $\overline{u}$ when below it. We note that because the optimal policy has the structure (\ref{thshd_pol}), we can restrict the search for $u_k$ in (\ref{DP_Eq}) to the set $\{\underline{u}$, $\overline{u}\}$ rather than the entire range of values between them and still preserve optimality. Furthermore, as we will see in the next section, this structure of an optimal policy can serve as the basis for simpler heuristics that could perform very close to the optimal. To get further intuition about the optimal policy, let us consider the case when the sequence of patient anatomy instances or costs $h(s_{k})$, $k = 1, 2, \ldots, N$, are known for the entire treatment. Then, it is clear that the solution would be to deliver $\overline{u}$ for the fractions with the smaller costs and $\underline{u}$ otherwise. Now, we can view our original problem, where the information about the patient anatomy is only available before the delivery of the fraction, as one of deciding whether the anatomy of any particular day will be one of the fractions with the smaller costs. The threshold $T_{k}(r_{k})$ then helps us make this determination. 

\subsection{Heuristic Policies based on Optimal Policy Structure}

Although it is possible to solve the problem exactly using the DP algorithm, we develop two heuristics that make use of the structure of an optimal policy and approximate the threshold $T_{k}(r_{k})$ in (\ref{thshd_pol}) by using: (i) the remaining dose $r_k$, which summarizes past information, and (ii) the distribution $p(\cdot)$, which summarizes information about the expected patient anatomy in the future. We believe such heuristics can provide simpler and intuitive solutions that can possibly be applied to more complex, high-dimensional problems, where using the DP algorithm is no longer computationally feasible. 

% "we let the threshold to 0" and "we let the threshold to 1" changed to "we set the threshold to 0" and "we set the threshold to 1"
Without loss of generality we assume $0 \leq h(s_k) \leq 1$, for all $s_k$. For simplicity we assume that $(\underline{u} + \overline{u})/2 = P/N$, so that $\underline{u}$ and $\overline{u}$ need each to be applied half of the time over the course of treatment. Our Heuristic $1$ which uses the following threshold:
\begin{equation}
T_{k}(r_{k}) = \left\{
\begin{array}{lll}
		0,				&		&\text{if $r_{k} = (N-k+1)\underline{u}$},\\
		1,			 	&		&\text{if $r_{k} = (N-k+1)\overline{u}$},\\
		M,                               &               &\text{otherwise},
	\end{array} \right.
	\label{thresh_heu1}
\end{equation}
where $M$ is the median of $h(s_k)$ which by definition satisfies $F^{\text{h}}(M) = \frac{1}{2}$. Such a policy has a simple interpretation: If the remaining dose $r_{k}$ in the $k$th fraction is such that we must deliver the smallest fraction size $\underline{u}$ for the remaining fractions (in which case $r_{k} = (N-k+1)\underline{u}$), we set the threshold to $0$, ensuring that regardless of the anatomy $s_{k}$, we always deliver the smallest fraction size $\underline{u}$. And, similarly, if the remaining dose $r_{k}$ is $(N-k+1)\overline{u}$, we set the threshold to $1$ and as a result, deliver the largest fraction size $\overline{u}$ for the remaining fractions. Otherwise, for the interesting case when $r_{k}$ is between $(N-k)\underline{u}$ and $(N-k)\overline{u}$, this policy simply delivers the smallest fraction size $\underline{u}$ when the cost $h(s_{k})$ is above its median $M$ (on average, this will happen half of the time) and the largest fraction size $\overline{u}$ when below it (on average, this will happen the other half of the time). Ignoring the possibility that the threshold $T_{k}(r_{k})$, for $k = 1, 2, \ldots, N$, can take extreme values (either $0$ or $1$), this policy is stationary, in the sense that the thresholds do not change with the fraction $k$. This is a simplistic approximation to the true values of $T_{k}(r_{k})$. The nice feature of this policy is that we do not need all of the information given in the probability distribution $p(\cdot)$; the only information required is the median $M$ of $h(s_k)$. This could be useful in a realistic setting in which we do not actually have accurate information about the distribution $p(\cdot)$. Here, one could estimate the median $M$ (perhaps by using statistical information from many patient datasets) and use the above heuristic policy.

An even better heuristic policy would likely use the entire distribution $p(\cdot)$ (as opposed to just a statistic such as the median or mean) to determine the threshold $T_k(r_k)$. Consider Heuristic $2$, which uses a threshold $T_k(r_k)$ which satisfies the following equation
\begin{equation}
F^{\text{h}}(T_k(r_k)) = \frac{i_k}{N-k+1},
\label{thresh_eq}
\end{equation}
where $i_k$ is the number of largest fraction sizes $\overline{u}$ left to deliver in the remaining $(N-k+1)$ fractions. Given that $T_k(r_k)$ is fixed for the remaining $(N-k+1)$ fractions, the left hand side of (\ref{thresh_eq}) represents the percentage of the remaining fractions for which we \emph{expect} to deliver the largest fraction size $\overline{u}$. And the right hand side represents the percentage of the remaining $(N-k+1)$ fractions for which we \emph{must} deliver the largest fraction size $\overline{u}$. In some sense, this threshold represents the best balance between what we expect to deliver and what we must deliver. For the uniform distribution, i.e. $h(s_k)\sim U[0,1]$, the threshold $T_k(r_k)$ for Heuristic $2$ has a simple closed form expression:
\begin{equation}
T_k(r_k) = \frac{i_k}{N-k+1}. \label{soph_thresh}
\end{equation}
Of course, it may not be possible to write $T_k(r_k)$ as a closed form expression for other (even common) distributions. However, for many of these distributions, the threshold that satisfies (\ref{thresh_eq}) can be evaluated by looking at tabulated values of the function $F^{\text{h}}(\cdot)$ (e.g., for the Gaussian distribution).\footnote{For the case of a discrete probability distribution with a few possible patient anatomy instances, a naive implementation of Heuristic $2$ can result in the threshold taking a value of $0$ even though it is not necessary to deliver the smallest fraction size $\underline{u}$ for the remaining fractions. In such cases, we forced the heuristic to deliver the largest fraction size $\overline{u}$ when $h(s_k) = 0$, and this resulted in better performance.}

\section{Results}

We discuss the results from implementing the adaptive algorithms (both exact and heuristic) in Matlab. For the problem parameters, we take the number of fractions $N$ to be $30$, the prescribed dose $P$ to be $60$ Gy, the smallest fraction size $\underline{u}$ to be $1.6$ Gy, the largest fraction size $\overline{u}$ to be $2.4$ Gy, the set of patient anatomy instances $S$ to be 10 equally spaced values between $0$ and $1$ representing the distance between the tumor and OAR (see Figure \ref{fig:dose_scaling_nice}), the distribution $p(\cdot)$ to be a discrete uniform, and the function $h(s_k)$ to be $1-s_k$. Essentially, we are allowing for a $20\%$ daily fraction size deviation from the standard $2$ Gy per fraction. 

\begin{figure}
\centerline{\includegraphics[width=130mm]{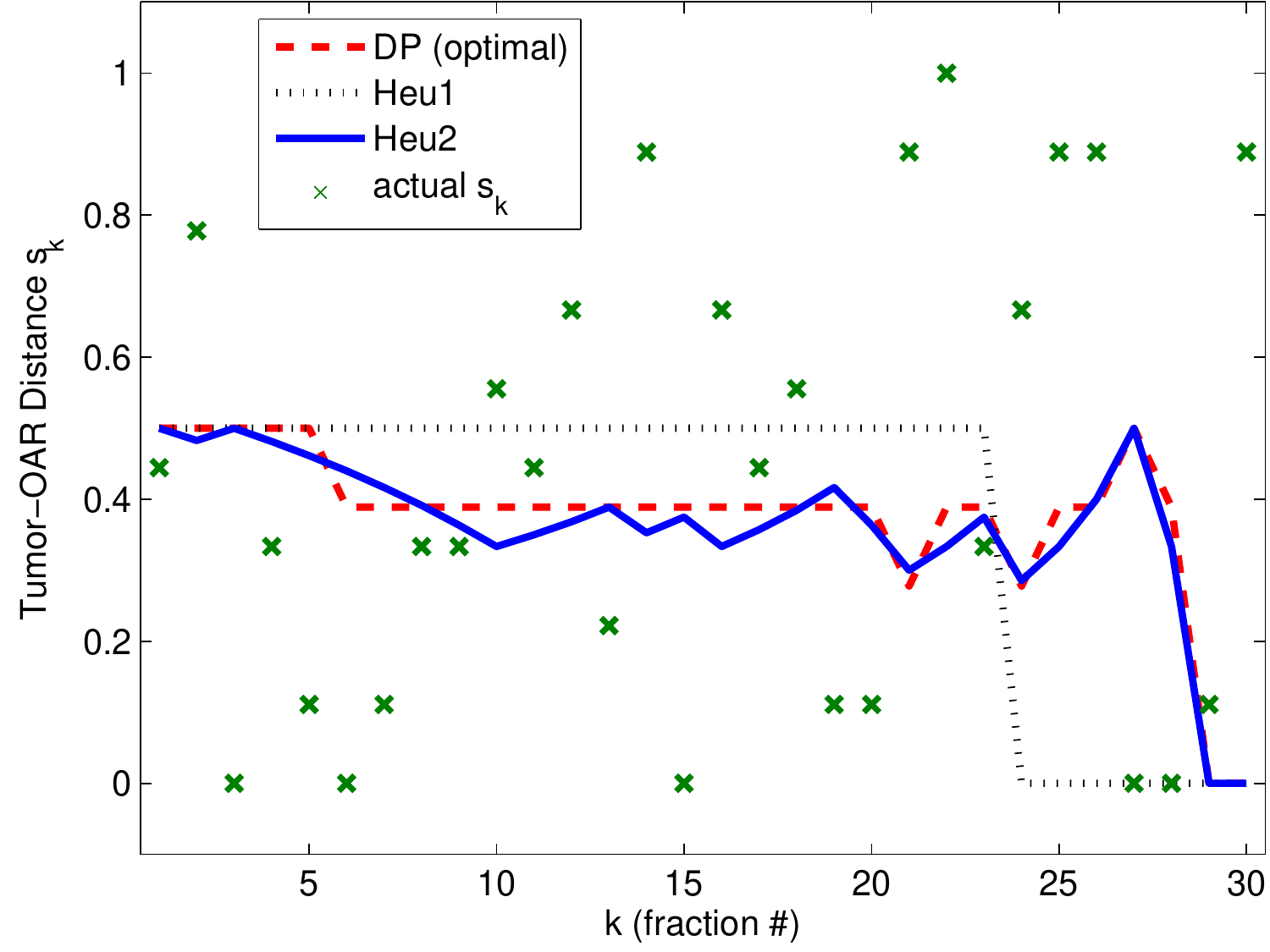}}
\caption{Thresholds of optimal and heuristic policies resulting from one treatment simulation run (i.e., one realization of the patient anatomy sequence $\{ s_1, s_2, \ldots, s_{N} \}$). For this 1-dimensional example, the threshold lines plotted represent the point at which a policy delivers the smallest fraction size $\underline{u}$ when $s_k$ is below it and the largest fraction size $\overline{u}$ when above it. These lines plotted are actually $1 - T_k(r_k)$ because we are plotting $s_k$ instead of $h(s_k) = 1 - s_k$. The `x' markers correspond to the actual realization of the tumor-OAR distance $s_k$. Heuristic $1$ (Heu1) makes a crude approximation to the optimal threshold while Heuristic $2$ (Heu2) follows it closely. Since the `x' markers are uniformly spread out and rarely take values near the thresholds, the heuristic algorithms perform well compared to the optimal DP approach. }
\label{fig:heuristics_vs_optimal}
\end{figure}

We find that both Heuristic $1$ and $2$ do well in approximating the optimal threshold, and as a result, perform numerically close to optimal. In Figure \ref{fig:heuristics_vs_optimal}, for one treatment simulation (i.e., one realization of the sequence $\{s_1, s_2, \ldots, s_{N} \}$), we show the thresholds of the optimal and heuristic policies. When the tumor-OAR distance $s_k$ is large and above the threshold, which indicates a favorable anatomy, the policy delivers the largest fraction size $\overline{u}$, and vice versa. We do note that, for this 1-dimensional setting, the threshold in Figure \ref{fig:heuristics_vs_optimal} is equal to $1-T_k(r_k)$ because we are plotting the tumor-OAR distance $s_k$ on the $y$-axis rather than $h(s_k)$. While Heuristic $2$ closely approximates the optimal threshold, Heuristic $1$ makes a crude approximation since it only uses the median $M$ of $h(s_k)$. Since the realized tumor-OAR distances $s_k$ (as shown by the `x' markers) are uniformly spread out and rarely take values near the thresholds, we see that the heuristic algorithms perform well. 

In Table \ref{tab:table2}, we see that when using a uniformly distributed motion model, the adaptive policies result in about a $10$\% decrease in dose to the OAR compared to that resulting from standard fractionation. Note that the DP approach represents the optimal policy, and hence, provides a baseline for comparison to the other heuristics. The difference in the dose to the OAR resulting from Heuristic $1$ and the DP approach is very little, which means that using a statistic such as the median $M$ of $h(s_k)$ is enough for achieving significant decrease in dose to the OAR. Such a policy could be advantageous in a realistic setting when it is not possible to have accurate information about the distribution $p(\cdot)$. As expected, Heuristic $2$ performs even better than Heuristic $1$ since it uses the entire distribution $p(\cdot)$ in the threshold computation. We can see that the numerical difference between the OAR dose resulting from Heuristic $2$ and the DP approach is not even visible when using two decimal places. Finally, we also simulate the algorithm in \cite{LCC2008} for comparison and notice that it is close to optimal as well, like the other heuristics. 

\begin{table}[t]
\caption{Using a uniformly distributed motion model and a 20\% daily fraction size deviation, we find about a 10\% decrease in dose to the OAR when using adaptive policies. The dose to the OAR is averaged over 10,000 treatment runs in order to report results to two decimal places. \\}
\centering
\begin{tabular}{|l|c|}
\hline
& Average Dose to OAR \\
\hline
Standard Fractionation & 30.00 \\
DP (Optimal) & 27.00 \\
Heuristic 1 & 27.13 \\
Heuristic 2 & 27.00 \\
Algorithm in \cite{LCC2008} & 27.07 \\
\hline
\end{tabular}
\label{tab:table2}
\end{table}

\begin{figure}
\centerline{\includegraphics[width=130mm]{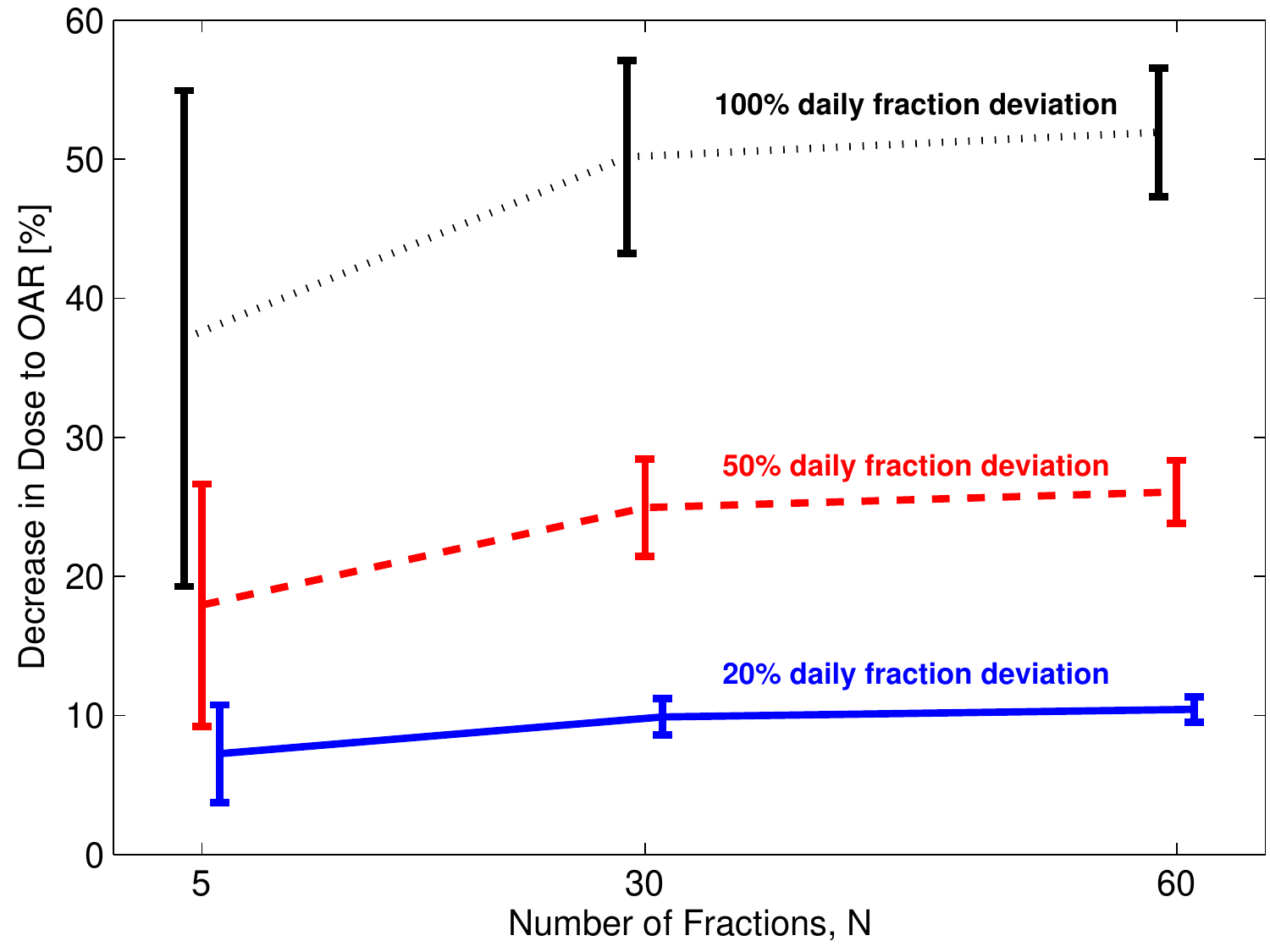}}
\caption{Comparing adaptive fractionation in hypo-, standard, and hyper-fractionated settings. We simulate the performance of just the optimal DP approach through 500 treatment runs. The fraction size is adjusted when varying the number of fractions $N$ so that the same prescribed dose $P$ is met at the end of treatment. The error bars correspond to one standard deviation, as estimated from the results of the 500 runs. We find a larger decrease in dose to the OAR when using more fractions and larger daily fraction size deviations.}
\label{fig:hypo_std_hyper}
\end{figure}

In Figure \ref{fig:hypo_std_hyper}, we vary both the number of fractions $N$ and the daily fraction size deviations, and simulate the decrease in the OAR dose when using the optimal DP approach. We use 20\%, 50\%, and 100\% daily fraction size deviations, and 5, 30, and 60 fractions for $N$. This allows us to understand the benefit of adaptive fractionation in hypo-, standard, and hyper-fractionation regimes. However, this may not be entirely accurate because we simply normalize the dose per fraction so that the same prescribed dose $P$ is met at the end of treatment, and we do not take into account the biological effect of varying $N$. The error bars in Figure \ref{fig:hypo_std_hyper} correspond to one standard deviation, as estimated from the simulation of 500 treatment runs. A larger number of fractions and daily fraction size deviation result in more chances to capitalize on favorable anatomy, and therefore, result in more gain. We see bigger error bars when using larger daily fraction size deviations due to the increased variation in the ability to capitalize on favorable anatomy. On the other hand, we see smaller error bars when increasing the number of fractions, which means the treatment outcome is more predictable; this is because with a large number of fractions, laws of large numbers (from probability theory) take effect. In summary, Figure \ref{fig:hypo_std_hyper} shows that the percentage decrease in the dose to the OAR varies significantly (anywhere from 5 - 55\%). But in general, we find more gain when using a larger number of fractions and daily fraction size deviations. 

\begin{figure}
\centerline{\includegraphics[width=170mm]{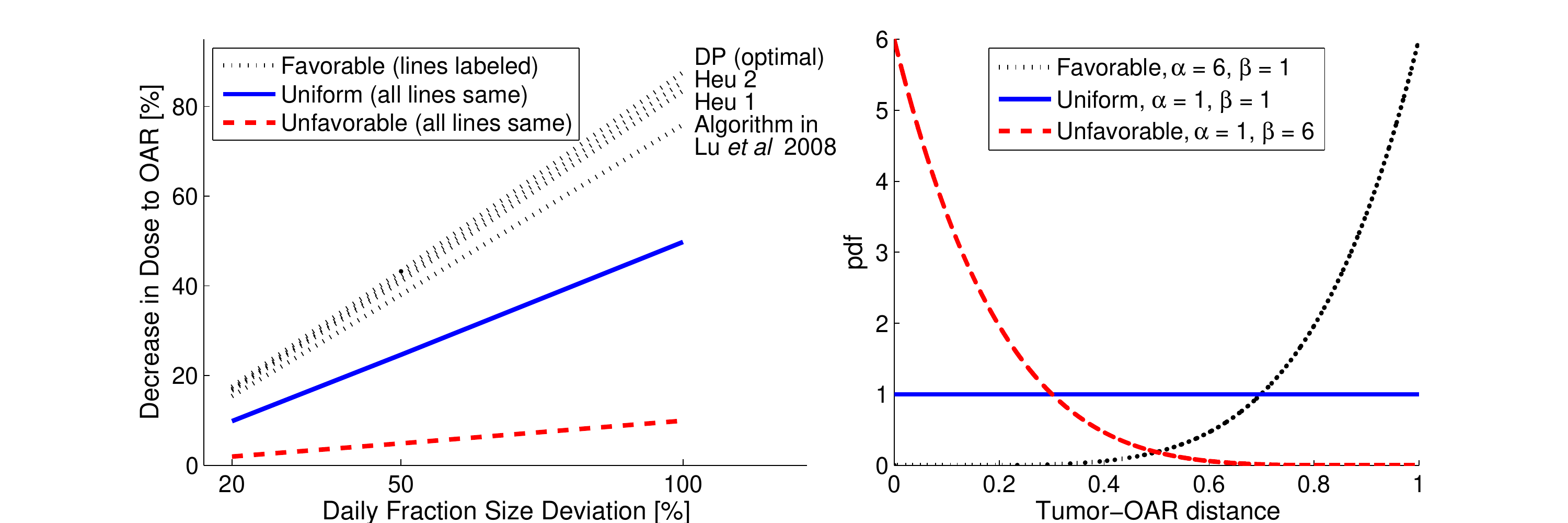}}
\caption{Results from varying the motion probability distribution. In the right panel, we show the three probability distributions used, each resulting from varying parameters of the beta distribution.  In the left panel, we show the average percentage decrease (from 500 treatment runs) in dose to the OAR for each of these probability distributions. For probability distributions in which the OAR tends to stay far away from the tumor, there is a larger decrease in dose to the OAR, and the optimal DP approach is at least 10\% better than the other heuristics.}
\label{fig:vary_prob_dist}
\end{figure}

As we see in Figure \ref{fig:vary_prob_dist}, the decrease in dose to the OAR is more pronounced when we have a high probability of large tumor-OAR distances. We use three distributions, each corresponding to parameters of the beta distribution ($p(s) = c \cdot s^{1-\alpha}(1-s)^{1-\beta}$, where $c$ is a normalization constant), and plot them in the right panel. As before, we use $10$ uniformly discretized values between $0$ and $1$ for the possible tumor-OAR distances. In the left panel, for the Unfavorable distribution, the percentage decrease in the OAR dose is minimal. On the other hand, there is a significant decrease in the OAR dose for the Uniform and Favorable distributions. We conclude that when the OAR tends to stay far away from the tumor, we see a larger decrease in dose to the OAR. In such cases, favorable anatomies are frequent enough so that the DP approach is able to make up for the small fraction sizes used for unfavorable anatomies. When allowing a 100\% daily fraction size deviations and using the Favorable distribution, there is at least a 10\% decrease in the OAR dose when comparing the optimal DP approach with the algorithm in \cite{LCC2008}. Here, the algorithm in \cite{LCC2008} ``believes" favorable anatomies will be frequent enough over the course of treatment so that no matter what fraction size is delivered in the initial fractions, it will be able to make up for the much more infrequent unfavorable anatomies.

\section{Discussion and Conclusions}

Realistic implementation of the approaches described in this paper require the following:
\begin{enumerate}
\item The motion probability distribution $p(\cdot)$ must be known. One could in principle collect and analyze population data, and assign probabilities corresponding to a few anatomy scenarios. However, this may not be an accurate representation of the patient-specific probability distribution. Updating the probability distribution as the sequence of patient anatomy instances are observed could be a topic of future research.
\item Immediately before the delivery of a fraction, imaging information about the patient anatomy must be available as well as reliable automatic contouring and/or contour registration algorithms to process this information. This would be needed in order to determine the favorability of the patient anatomy in any particular day. We can be optimistic that such technology will be available in the near future.
\item The OAR dose per unit dose to the tumor, $h(s_k)$, must be computable before the delivery of each fraction. In order to speed up the computation, one may use the same dose decomposition matrix from the initial CT scan as an approximation and determine the dose projected on the OAR in the new CT. We emphasize that this model does not depend on using the OAR dose per unit tumor dose for $h(s_k)$; we simply need a function $h(s_k)$ which tells us how favorable a particular anatomy $s_k$ is.
\end{enumerate}

\noindent Other assumptions of our model include the following:
\begin{enumerate}
\item One primary OAR is the basis for making decisions about the fraction size. Multiple OARs can be used, for example, by using for $h(s_k)$ a weighted combination of the dose to each OAR per unit dose to the tumor. However, this does not capture the true tradeoff between the various OARs because there is also generally an upper limit to the dose of each OAR. The model would better represent the radiation therapy problem if, for example, the non-linear NTCP curves were incorporated into the cost function. There is a potential for further research work here.
\item The prescribed dose to the tumor must be met exactly and is penalized with an infinite cost. One may also consider using a penalty (e.g., quadratic) on the deviation from the prescribed dose, which would possibly result in a ``smoother" optimal policy that does not pick the extreme fraction sizes.
% added this item discussing random vs. systematic changes.
\item Variations in the patient anatomy are dose-independent and random. In order to incorporate systematic time varying trends (e.g., tumor shrinkage), we can re-optimize the treatment plan (the dose profile) midway through treatment. Then, we can use one of the adaptive fractionation methods in this paper for the first half of treatment assuming no systematic changes. For the latter half of treatment, we would re-optimize the treatment plan, update the motion probability distribution and/or escalate the prescribed dose if necessary, and restart the adaptive fractionation method. For other systematic offsets (e.g, patient setup errors), we assume daily imaging modalities are accurate enough for correction.
\item When the daily fraction size deviations are not too large, the biological impact of a varied fractionation scheme is negligible. It is likely that deviating 20\% from standard fractionation does not result in a major difference between physical dose and biological dose \cite{BoP2006}. Dose deviations of 50\% and 100\%, however, need further study. Using large deviations and varied fractionation may require additional considerations such as changes in the onset of early and late reactions. A biologically based adaptive fractionation approach is given in \cite{CLC2008}.
\end{enumerate}

Under our framework, it is possible to derive the algorithm in \cite{LCC2008} and see that it is a variant of the approximate DP approach known as the open-loop feedback control (refer to \cite{Ber2007} for a description of such an approach). From the numerical results, we conclude that the algorithm in \cite{LCC2008} performs very close to optimal for almost all cases. However, we do see that the DP approach performs about 10\% better when allowing 100\% daily fraction size deviations and using a probability distribution that favors large tumor-OAR distances. One difference in the way these algorithms would be used in practice is that the DP approach involves using a table lookup on-line, while the algorithm in \cite{LCC2008} requires solving a linear programming (LP) problem right before the delivery of every fraction. Though a table lookup is quicker, solving a LP for this problem, where we are simply searching for a scalar variable $u_k$, is also very fast and can be done before each fraction without much time overhead.

% previous: Adaptive fractionation will, therefore, lead to a more uniform dose to the OAR fraction to fraction even though we are not taking this into account in the optimization. We believe there is potential for future work along these lines which compares the biological benefit of varying the fraction size.
Based on the linear-quadratic model of radiation effects, varying the fraction size during the course of treatment while ensuring a fixed total prescribed dose leads to a higher TCP \cite{BoP2006}. One might argue that this benefit is neutralized by an increase in the NTCP for the OAR. However, assuming that the OAR has an underlying motion, even standard fractionation results in a different dose to the OAR from fraction to fraction. Adaptive fractionation will lead to a more uniform dose to the OAR because a small fraction size is delivered when the OAR-to-tumor dose ratio $h(s_k)$ is large (i.e., when the tumor and OAR are close together), and vice versa. Further research work can be done to evaluate the biological benefit of varying the fraction size.

We have posed the adaptive fractionation problem in a theoretical framework and have provided several solution methods. First, we used the DP algorithm to establish a benchmark and to solve the problem exactly. This allowed us to show that the simple heuristics proposed in this paper were numerically near-optimal. One of these heuristics only uses a statistic, such as the median or mean, rather than the entire probability distribution. Such a policy can provide a quick way to estimate the best possible benefit of using an adaptive fraction size in a realistic setting. We have demonstrated through numerical simulations that we can expect a significant decrease in dose to the OAR when: (i) we have a high probability of large tumor-OAR distances, (ii) we use many fractions (as in a hyper-fractionated setting), and (iii) we allow large daily fraction size deviations. We expect adaptive fractionation to be beneficial for disease sites in which the OAR exhibits significant motion from day to day. Some examples include pelvic cases such as rectal \cite{NRY2004}, prostate \cite{WES2011}, and cervical \cite{GWA2011} cancers. Future work will include determining which disease sites will benefit from adaptive fractionation and demonstrating the gain for these sites.

\ack
This work was partially supported by Siemens. %Thanks to ... for helpful suggestions.

\appendix
\section*{Appendix. Theoretical Results and Proofs}
\setcounter{section}{1}

% added footnote
We provide some additional details about the theoretical results in this appendix. We note that although we assumed $s_k$ to be independent and identically distributed, we can generalize the results to the case of correlated motion using the same arguments below.\footnote{To be precise, our methods apply to the case where the sequence of patient anatomy instances satisfy the Markov property. That is, the patient anatomy $s_{k+1}$ is only dependent on $s_k$ and not on previous anatomies before the $k$th day. In this case, the dependencies would be summarized in a new probability distribution $p_k(s_{k+1} | s_k)$ and the same analysis goes through.} To facilitate the discussion, we define 
\begin{equation}
\underline{B}_k(r_k) = \text{max}(\underline{u}, r_{k}-(N-k)\overline{u})
\end{equation}
and
\begin{equation}
\overline{B}_k(r_k) = \text{min}(\overline{u}, r_{k}-(N-k)\underline{u})),
\end{equation}
which are the minimum and maximum allowable fraction sizes, respectively, in the $k$th fraction. We can verify this by noting that because the prescribed dose to the tumor must be met exactly, the remaining dose $r_{k}$ must be between the smallest and the largest possible fraction size deliverable to the tumor for the remaining fractions (i.e., $(N-k+1)\underline{u}$ and $(N-k+1)\overline{u}$). Now we can rewrite the DP equation as
\begin{equation}
J_{k}(r_{k}, s_{k}) = \underset{\underline{B}_k(r_k) \leq u_{k} \leq \overline{B}_k(r_k)}{\text{min}} \Big( u_{k}h(s_{k}) + \mathbb{E} \left[ J_{k+1}(r_k-u_k, s_{k+1}) \right] \Big), \label{DP_Eq2}
\end{equation}
with the same terminal condition as before. We describe the qualitative structure of the cost-to-go function in the following theorem.

\vspace{0.2cm}

\noindent \textbf{Theorem. }\emph{The cost-to-go function $J_{k}(r_{k}, s_{k})$, for $k = 1, 2, \ldots, N$, is piecewise linear, continuous, non-decreasing, and convex in $r_{k}$ for feasible $r_{k}$, with breakpoints at $(N-k+1-i)\underline{u} + i\overline{u}$, for $i = 0, 1, \ldots, N-k+1$.}

\vspace{0.2cm}

\noindent\emph{Proof Sketch. }Here, we give only a proof sketch of the piecewise linearity of the cost-to-go function, which is the key property that gives rise to the special structure of an optimal policy. We proceed by induction. For the base case, let $k = N$. It is clear that the optimal decision in this last fraction, $\mu_{N}^*(r_{N}, s_{N})$, is equal to the remaining dose $r_{N}$. (Otherwise, we would incur an infinite penalty for not meeting the prescribed dose exactly.) In this case, $J_{N}(r_{N}, s_{N})$ has the desired piecewise linear form (in fact, it is simply equal to the linear function $r_{N}h(s_{N})$) over the range of feasible $r_N$. Now, assume $J_{k+1}(r_{k+1}, s_{k+1})$ has the desired piecewise linear form given in the statement of the theorem. Then, in (\ref{DP_Eq2}), we are minimizing a piecewise linear function with breakpoints separated by $(\overline{u}-\underline{u})$ over an interval $\left[ \underline{B}_k(r_k), \overline{B}_k(r_k) \right]$, which is of length at most $(\overline{u}-\underline{u})$. It follows that one of three possibilities attains the minimum: $\underline{B}_k(r_k)$, $r_k-((N-k+1-i^*)\underline{u} + i^*\overline{u})$, or $\overline{B}_k(r_k)$, where $i^*$ is an integer between $0$ and $N-k+1$. We can substitute each of these possibilities in (\ref{DP_Eq2}), and after some algebra, show that $J_k(r_k, s_k)$ has the desired piecewise linear form. This completes the induction. \hfill$\square$ 

\vspace{0.2cm}

The consequence of the above theorem is that when it is possible to deliver the treatment with a sequence of smallest and largest fraction sizes (i.e., when $r_k$ is a nonnegative integer combination of $\underline{u}$ and $\overline{u}$), an optimal policy does exactly that. This is stated mathematically in the following corollary.

\vspace{0.2cm}

\noindent \textbf{Corollary. }\emph{If there exists an integer $i$ between $0$ and $N$ such that the initial remaining dose (or the prescribed dose) can be written as $r_{1} = (N-i)\underline{u} + i\overline{u}$, then an optimal policy has a threshold form:
\begin{equation}
	\mu_{k}^*(r_{k}, s_{k}) = \left\{
	\begin{array}{lll}
		\underline{u},		&	&\text{if $h(s_{k}) \geq T_{k}(r_{k})$,}\\
		\overline{u},		&	&\text{if $h(s_{k}) < T_{k}(r_{k})$,}\\
	\end{array} \right. 
\end{equation}
for $k = 1, 2, \ldots, N$.}

\vspace{0.2cm}

\noindent\emph{Proof. }We proceed by induction. Let $k = 1$ and assume, for the base case, that there exists an integer $i_1$ between $0$ and $N$ such that the initial remaining dose (or the prescribed dose) can be written as $r_{1} = (N-i_1)\underline{u} + i_1\overline{u}$. We consider three cases:
\begin{enumerate}
\item When $i_1 = 0$, i.e. $r_1 = N\underline{u}$, the only possible solution is $u_{1} = u_{2} = \ldots = u_{N} = \underline{u}$, in which case we are done.
\item Similarly, when $i_1 = N$, $u_{1} = u_{2} = \ldots = u_{N} = \overline{u}$ is the only solution.
\item We consider the more interesting case when $i_1$ is an integer between $1$ and $N-1$. First, we notice that for any choice of $u_1$ between $\underline{u}$ and $\overline{u}$, $r_2 = r_1-u_1$ remains feasible, and hence, the cost-to-go function $J_{2}(r_1-u_1,s)$ is finite for all anatomy instances $s$. Second, since $r_{1} = (N-i_1)\underline{u} + i_1\overline{u}$, from the above theorem, we have that $J_2(r_1-u_1, s)$ is linear in $u_1$ for all values in between $\underline{u}$ and $\overline{u}$. Since taking an expectation preserves linearity, the function 
$\mathbb{E} \left[ J_{2}(r_1-u_1, s_{2}) \right] $
 is also linear in $u_1$. Now, in the DP equation
\begin{equation}
J_{1}(r_{1}, s_{1}) = \underset{\underline{u} \leq u_{1} \leq \overline{u}}{\text{min}} \Big( u_{1}h(s_{1}) + \mathbb{E} \left[ J_{2}(r_1-u_1, s_{2}) \right]\Big),
\label{DP_coro}
\end{equation}
we are minimizing a linear function because adding a linear function $u_1h(s_1)$ preserves linearity. So, for any feasible $r_1$, we can write
\begin{equation}
\mathbb{E} \left[ J_{2}(r_1-u_1, s_{2}) \right]= a(r_1)u_1 + b(r_1),
\end{equation}
where $a(r_1)$ and $b(r_1)$ represent the slope and intercept, respectively. Let $T_1(r_1) = -a(r_1)$. Then, since we are minimizing a linear function over an interval in (\ref{DP_coro}), $\mu_1^*(r_1, s_1)$ has the desired threshold form:
\begin{equation}
	\mu_{1}^*(r_{1}, s_{1}) = \left\{
	\begin{array}{lll}
		\underline{u},		&	&\text{if $h(s_{1}) \geq T_{1}(r_{1})$,}\\
		\overline{u},		&	&\text{if $h(s_{1}) < T_{1}(r_{1})$.}\\
	\end{array} \right. 
\end{equation}
Now, it is clear that from the above form for $\mu_{1}^*(r_{1}, s_{1})$, there exists an integer $i_2$ between $0$ and $N-1$ such that the remaining dose in the next fraction can be written as $r_{2} = (N-1-i_2)\underline{u} + i_2\overline{u}$. We can complete the induction by assuming the appropriate form for $r_k$ in the induction hypothesis and following the same line of argument as above. \hfill$\square$
\end{enumerate}

The assumption that the initial remaining dose can be written as $r_1 = (N-i)\underline{u} + i\overline{u}$ simply requires that it be possible to deliver the treatment with a sequence of smallest fraction sizes $\underline{u}$ and largest ones $\overline{u}$. Otherwise, it would not be possible for the cumulative sum of the fraction sizes to be equal to the prescribed dose when restricting to only the smallest or the largest fraction size. In that respect, the assumption here is reasonable. Though there are generalizations to the structure of an optimal policy when this assumption is not satisfied, it is not necessary to discuss them further since this assumption is generally satisfied in a realistic setting or at least satisfiable with a slight modification of the lower bound $\underline{u}$ and upper bound $\overline{u}$.

\bigskip
\section*{References}
\bibliographystyle{abbrv}
\bibliography{ref}

\newcommand{\noopsort}[1]{} \newcommand{\printfirst}[2]{#1}
  \newcommand{\singleletter}[1]{#1} \newcommand{\switchargs}[2]{#2#1}
\begin{thebibliography}{10}

\bibitem{Ber2007}
D.~P. Bertsekas.
\newblock {\em Dynamic Programming and Optimal Control}.
\newblock Athena Scientific, 2007.

\bibitem{BoP2006}
T.~Bortfeld and H.~Paganetti.
\newblock The biologic relevance of daily dose variations in adaptive treatment
  planning.
\newblock {\em Int. J. Radiation Oncology Biol. Phys.}, 65:899--906, 2006.

\bibitem{CLC2008}
M.~Chen, W.~Lu, Q.~Chen, K.~Ruchala, and G.~Olivera.
\newblock Adaptive fractionation therapy: {II}. {B}iological effective dose.
\newblock {\em Phys. Med. Biol.}, 53:5513--5525, 2008.

\bibitem{CDL2005}
L.~E. Court, L.~Dong, A.~K. Lee, R.~Cheung, M.~D. Bonnen, J.~O'Daniel, H.~Wang,
  R.~Mohan, and D.~Kuban.
\newblock An automatic {CT}-guided adaptive radiation therapy technique by
  online modification of multileaf collimator leaf positions for prostate
  cancer.
\newblock {\em Int. J. Radiation Oncology Biol. Phys.}, 62:154--163, 2005.

\bibitem{ZAX2007}
A.~de~la Zerda, B.~Armbruster, and L.~Xing.
\newblock Formulating adaptive radiation therapy ({ART}) treatment planning
  into a closed-loop control framework.
\newblock {\em Phys. Med. Biol.}, 52:4137--4153, 2007.

\bibitem{DeF2008}
G.~Deng and M.~C. Ferris.
\newblock Neuro-dynamic programming for fractionated radiotherapy planning.
\newblock {\em Springer Optimization and Its Applications}, 12:47--70, 2008.

\bibitem{FeV2004}
M.~C. Ferris and M.~M. Voelker.
\newblock Fractionation in radiation treatment planning.
\newblock {\em Math. Program.}, 101:387--413, 2004.

\bibitem{GWA2011}
J.~J. Gordon, E.~Weiss, O.~K. Abayomi, J.~V. Siebers, and N.~Dogan.
\newblock The effect of uterine motion and uterine margins on target and normal
  tissue doses in intensity modulated radiation therapy of cervical cancer.
\newblock {\em Phys. Med. Biol.}, 56:2887--2901, 2011.

\bibitem{GWR2011}
M.~Guckenberger, J.~Wilbert, A.~Richter, K.~Baier, and M.~Flentje.
\newblock Potential of adaptive radiotherapy to escalate the radiation dose in
  combined radiochemotherapy for locally advanced non–small cell lung cancer.
\newblock {\em Int. J. Radiation Oncology Biol. Phys.}, 2011.

\bibitem{Lan2011}
K.~M. Langen.
\newblock Uncertainties and limitations in adaptive radiotherapy.
\newblock In J.~R. Palta and T.~R. Mackie, editors, {\em Uncertainties in
  external beam radiation therapy}, pages 443--470. Medical Physics Publishing,
  2011.

\bibitem{LCC2008}
W.~Lu, M.~Chen, Q.~Chen, K.~Ruchala, and G.~Olivera.
\newblock Adaptive fractionation therapy: {I}. {B}asic concept and strategy.
\newblock {\em Phys. Med. Biol.}, 53:5495--5511, 2008.

\bibitem{MZW2005}
R.~Mohan, X.~Zhang, H.~Wang, Y.~Kang, X.~Wang, H.~Liu, K.~K. Ang, D.~Kuban, and
  L.~Dong.
\newblock Use of deformed intensity distributions for on-line modification of
  image-guided {IMRT} to account for interfractional anatomic changes.
\newblock {\em Int. J. Radiation Oncology Biol. Phys.}, 61:1258--1266, 2005.

\bibitem{NRY2004}
J.~J. Nuyttens, J.~M. Robertson, D.~Yan, and A.~Martinez.
\newblock The influence of small bowel motion on both a conventional
  three-field and intensity modulated radiation therapy (imrt) for rectal
  cancer.
\newblock {\em Cancer/Radioth\'erapie}, 8:297--304, 2004.

\bibitem{PTE2010}
G.~Poludniowski, M.~D.~R. Thomas, P.~M. Evans, and S.~Webb.
\newblock {CT} reconstruction from portal images acquired during
  volumetric-modulated arc therapy.
\newblock {\em Phys. Med. Biol.}, 55:5635--5651, 2010.

\bibitem{RFL2004}
H.~Rehbinder, C.~Forsgren, and J.~L{\"{o}}f.
\newblock Adaptive radiation therapy for compensation of errors in patient
  setup and treatment delivery.
\newblock {\em Med. Phys.}, 31:3363--3371, 2004.

\bibitem{SEP2010}
M.~Y. Sir, M.~A. Epelman, and S.~M. Pollock.
\newblock Stochastic programming for off-line adaptive radiotherapy.
\newblock {\em Annals of Operations Research}, 179, 2010.

\bibitem{WES2011}
Y.~Wang, J.~A. Efstathiou, G.~C. Sharp, H.-M. Lu, I.~F. Ciernik, and A.~V.
  Trofimov.
\newblock Evaluation of the dosimetric impact of interfractional anatomical
  variations on prostate proton therapy using daily in-room ct images.
\newblock {\em Med. Phys.}, 38:4623--4633, 2011.

\bibitem{Web2008a}
S.~Webb.
\newblock Adapting {IMRT} delivery fraction-by-fraction to cater for variable
  intrafraction motion.
\newblock {\em Phys. Med. Biol.}, 53:1--21, 2008.

\bibitem{WeB2008b}
S.~Webb and T.~Bortfeld.
\newblock A new way of adapting {IMRT} delivery fraction-by-fraction to cater
  for variable intrafraction motion.
\newblock {\em Phys. Med. Biol.}, 53:5177--5191, 2008.

\bibitem{WTW2008}
Q.~J. Wu, D.~Thongphiew, Z.~Wang, B.~Mathayomchan, V.~Chankong, S.~Yoo, W.~R.
  Lee, and F.-F. Yin.
\newblock On-line re-optimization of prostate {IMRT} plans for adaptive
  radiation therapy.
\newblock {\em Phys. Med. Biol.}, 53:673--691, 2008.

\bibitem{YLB2000}
D.~Yan, D.~Lockman, D.~Brabbins, L.~Tyburski, and A.~Martinez.
\newblock An off-line strategy for constructing a patient-specific planning
  target volume in adaptive treatment process for prostate cancer.
\newblock {\em Int. J. Radiation Oncology Biol. Phys.}, 48:289--302, 2000.

\bibitem{YVW1997}
D.~Yan, F.~Vicini, J.~Wong, and A.~Martinez.
\newblock Adaptive radiation therapy.
\newblock {\em Phys. Med. Biol.}, 42:123--132, 1997.

\bibitem{YZJ1998}
D.~Yan, E.~Ziaja, D.~Jaffray, J.~Wong, D.~Brabbins, F.~Vicini, and A.~Martinez.
\newblock The use of adaptive radiation therapy to reduce setup error: {A}
  prospective clinical study.
\newblock {\em Int. J. Radiation Oncology Biol. Phys.}, 41:715--720, 1998.

\end{thebibliography}
\end{document}